\documentclass[aps,prc,twocolumn,floatfix,showpacs,
nofootinbib,amsmath,amssymb]{revtex4-1}
\usepackage{graphicx}
\usepackage{dcolumn}
\usepackage{bm}
\usepackage{color}

\newcommand{\be}{\begin{equation}}
\newcommand{\ee}{\end{equation}}
\newcommand{\ba}{\begin{eqnarray}}
\newcommand{\ea}{\end{eqnarray}}
\newcommand{\bd}{\begin{displaymath}}
\newcommand{\ed}{\end{displaymath}}
\newcommand{\bea}{\begin{eqnarray}}
\newcommand{\eea}{\end{eqnarray}}

\renewcommand{\vec}[1]{\mbox{\boldmath$#1$}}

\def\thalf{{\textstyle{\frac{1}{2}}}}

\def\oneth{{\textstyle{\frac{1}{3}}}}
\def\twoth{{\textstyle{\frac{2}{3}}}}
\def\oneqt{{\textstyle{\frac{1}{4}}}}

\begin{document}
\title{$\Lambda$ and $\bar{\Lambda}$ spin interaction with meson fields generated by the baryon current in high energy nuclear collisions}

\author{L. P. Csernai$^1$, J. I. Kapusta$^2$ and T. Welle$^2$}
\smallskip

\affiliation{$^1$Institute of Physics and Technology, University of Bergen,
Allegaten 55, 5007 Bergen, Norway \\
$^2$School of Physics and Astronomy, University of Minnesota, Minneapolis, Minnesota 55455 USA}

\begin{abstract}
We propose a dynamical mechanism which provides an interaction between the spins of hyperons and antihyperons and the vorticity of the baryon current in noncentral high energy nuclear collisions.  The interaction is mediated by massive vector and scalar bosons which is well known to describe the nuclear spin-orbit force.  It follows from the Foldy-Wouthuysen transformation and leads to a strong-interaction Zeeman effect.  The interaction may explain the difference in polarizations of $\Lambda$ and $\bar{\Lambda}$ hyperons as measured by the STAR Collaboration at the BNL Relativistic Heavy Ion Collider.  The signs and magnitudes of the meson-baryon couplings are closely connected to the binding energies of hypernuclei and to the abundance of hyperons in neutron stars.
\end{abstract}
\date{\today}

\maketitle 


Experiments at the BNL Relativistic Heavy Ion Collider (RHIC) and at the CERN Large Hadron Collider (LHC) have provided a wealth of data on the hot and dense matter created in collisions between heavy ions \cite{QMseries}.  Among these data are the coefficients of a Fourier expansion in the azimuthal angle for various physical observables.  They provide strong evidence for collective expansion of the hot and dense matter and provide information on transport coefficients such as the shear viscosity \cite{whitepaper}.  In addition, the polarization of $\Lambda$ and $\bar{\Lambda}$ hyperons was proposed as yet another observable that provides information on collective flow, in particular vorticity \cite{Wang1,Becattini1}.  Measurements of the polarizations have been made by the STAR Collaboration from the lowest to the highest beam energies at RHIC \cite{FirstSTAR,Nature,SecondSTAR}, noting that RHIC produces matter with the highest vorticity ever observed.

The standard picture of $\Lambda$ and $\bar{\Lambda}$ polarization in non-central heavy ion collisions assumes equipartition of energy \cite{Becattini2,Becattini3}.  But there is a potential puzzle presented by the experimental data: The $\bar{\Lambda}$ polarization is greater than the $\Lambda$ polarization by a factor of 4 at the level of two standard deviations at $\sqrt{s_{NN}} = 7.7$ GeV for Au+Au collisions.  Both the difference between the two and their absolute values decrease with increasing beam energy until they are approximately equal at $\sqrt{s_{NN}} = 200$ GeV, albeit only at the one-standard-deviation level, whereas equipartition would suggest no difference.  The interaction that we propose addresses the issue of the polarization difference.

It has been known since the early days of the nuclear shell model that a spin-orbit interaction is required to explain the single-particle energy levels \cite{nuclearSpinOrbit}.  It was subsequently shown that attractive scalar and repulsive vector meson exchanges naturally lead to such spin-orbit interactions via a nonrelativistic reduction of the Dirac equation \cite{Duerr}.  Starting with the so-called Walecka model \cite{Walecka74} much success has been achieved in describing nuclear structure, proton-nucleus scattering, and high density matter using various versions of these relativistic Lagrangians incorporating baryons and mesons \cite{Serot-Walecka,Kapusta-Gale}.  The fact that they include the strong interaction equivalent of the magnetic force and the spin-orbit force, including hyperons \cite{Dover-Gal}, suggests that this approach provides a natural explanation for the interaction between spin and vorticity and for the difference between $\Lambda$ and $\bar{\Lambda}$ polarizations.

Suppose that the strong interaction among bayons is mediated by a scalar field $\sigma$ and a vector field $V^{\mu}$.  The effective Lagrangian is
\ba
\lefteqn{{\cal L}_{\rm eff} = \sum_j \bar{\psi}_j ( i \! \not\!\partial - m_{j} 
+ g_{\sigma j} \sigma - g_{V j} \not\!V ) \psi_j} \nonumber \\
&& + \thalf \left( \partial_{\mu} \sigma \partial^{\mu} \sigma
- m_{\sigma}^2 \sigma^2 \right)
 - \oneqt V^{\mu\nu}V_{\mu\nu}
+ \thalf m_{V}^2 V_{\mu}V^{\mu} \,.
\label{Leff}
\ea
Here $j$ represents one of the spin-1/2 baryons in the octet, and the field strength tensor for the vector field is  
\be
V_{\mu\nu} = \partial_\mu V_\nu - \partial_\nu  V_\mu \,.
\label{e1}
\ee
In general there may be a potential $U(\sigma)$ which has terms cubic and quartic in $\sigma$ but its exact form will not be needed here.

One may perform a Foldy-Wouthuysen transformation \cite{FW,BD,Greiner} (an expansion in powers of the inverse of the baryon mass; higher order corrections may be found in \cite{Hinschberger}) to obtain the nonrelativistic interaction between the vector field and the spin operator $\vec{S}$ of the $\Lambda$ and $\bar{\Lambda}$.  (We set $\hbar = c = 1$.) The interaction of the spin with the vector meson is
\be
H^V_{\rm spin} =
- \frac{g_{V\Lambda}}{m_\Lambda} \beta \
\vec{S} \cdot \vec{B}_V 
- i \frac{g_{V\Lambda}}{4 m^2_\Lambda}
\vec{S} \cdot \vec{\nabla}\times  \vec{E}_V 
- \frac{g_{V\Lambda}}{2 m^2_\Lambda}
\vec{S} \cdot \vec{E}_V \times \vec{p}
\label{e2}
\ee
where 
$\vec{E}_V$ and  $\vec{B}_V$ are the vector meson electric and magnetic fields corresponding to Eq. (\ref{e1}), 
$\vec{p}$ is the momentum of the $\Lambda$ or $\bar{\Lambda}$, and 
\be
\beta = 
\left( \begin{array}{cc}
		1 & 0 \\ 0 & -1
\end{array} \right)
\label{e3}
\ee
is the usual Dirac $4 \times 4$ $\beta$ matrix.  When acting on the spinors of $\Lambda$ and $\bar{\Lambda}$ they result in opposite signs whereas the second and third terms have the same sign.  The second and third terms contribute to the usual nuclear spin-orbit energy.  Only their sum is Hermitian, not the individual terms.  (According to the Bianchi identity we can replace $\vec{\nabla}\times  \vec{E}_V$ with $-\partial \vec{B}_V/\partial t$.)  For a spherically symmetric static potential only the third term remains, which becomes
\be
H^V_{\rm spin-orbit} = \frac{g_{V\Lambda}}{2 m^2_\Lambda}
\frac{1}{r}\frac{\partial V_0}{\partial r} \vec{S} \cdot \vec{L}
\label{e4}
\ee
where $\vec{L} = \vec{r}\times\vec{p}$ is the orbital angular momentum.

For the scalar field the spin-orbit interaction is
\be
H^\sigma_{\rm spin-orbit} = \frac{g_{\sigma\Lambda}}{2 m^2_\Lambda} \vec{S} \cdot \vec{\nabla} \sigma \times \vec{p}
\label{e5}
\ee
while there is no ``magnetic" interaction.  For central potentials this becomes
\be
H^\sigma_{\rm spin-orbit} = \frac{g_{\sigma\Lambda}}{2 m^2_\Lambda}
\frac{1}{r}\frac{\partial \sigma}{\partial r} \vec{S} \cdot \vec{L} \,.
\label{e6}
\ee
In atomic nuclei $V$ is identified with the $\omega$ vector meson.  A survey of results in the literature leads to $g_{\omega N} \approx 8.646$ and $g_{\sigma N} \approx 8.685$ \cite{Kapusta-Gale}.  With the sign convention used here $\sigma > 0$ represents an attractive interaction and $\omega_0 > 0$ represents a repulsive interaction.  They contribute with the same sign to the spin-orbit interaction with approximately equal strengths, whereas their contributions to the total binding energy approximately cancel.

In the mean field approximation the vector field is calculated as follows \cite{Walecka74,Serot-Walecka,Kapusta-Gale}:
\be
\partial_\mu V^{\mu\nu} + m^2_{V} V^\nu = \sum_j g_{V j} J^\nu_j \,.
\label{field_omega}
\ee
Here $J^\mu_j$ is the baryon current $\langle \bar{\psi} \gamma^\mu \psi\rangle$ contributed by species $j$, such that protons and antiprotons contribute with opposite signs, for example.  The mean scalar field is determined by
\be
\partial^2 \sigma + m_{\sigma}^2 \sigma + \frac{dU}{d\sigma} = \sum_j g_{\sigma j} n_{sj}
\label{field_sigma}
\ee
where $n_{sj}$ is the scalar density $\langle \bar{\psi} \psi\rangle$ contributed by species $j$, such that protons and antiprotons contribute with the same sign, for example.
These interactions are anticipated to become relevant around the time of hadronization of the hot and dense matter created in the collisions which is generally accepted to be on the order of 3 to 5 fm/c or longer.  The corresponding energy scale is much less than $m_\omega = 783$ MeV and $m_\sigma \approx 550$ MeV so that the derivatives in Eqs. (\ref{field_omega}) and (\ref{field_sigma}) can be neglected.

For noncentral potentials, $\vec{\nabla} \times \vec{E_V} = -\partial \vec{B}_V/\partial t \ne 0$, $\vec{\nabla} \sigma \ne (\vec{r}/r^2)\partial \sigma/\partial r$, the spin-orbit terms represent an exchange of energy and angular momentum with the fields.  For some systems in the condensed matter context, the electromagnetic spin-orbit interaction has been used to derive the Gilbert term which describes Gilbert damping, the rate at which magnetization relaxes to equilibrium, in Refs. \cite{Hickey,Mondal}.  The damping of magnetization is commensurate with the emission of electromagnetic radiation.  Assuming that the baryons in high energy nuclear collisions have a vortical flow motion, the scalar and vector meson interactions given above can provide a mechanism for hyperon polarization.  In addition, note that the ``magnetic" interaction is opposite in sign for hyperons and antihyperons due to the factor of $\beta$.

We can make a simple estimate of the magnitudes and signs of the effects.  We work in the center-of-momentum frame of the colliding nuclei at mid-rapidity and neglect Lorentz $\gamma$ factors.  The $x$-$z$ plane is taken as the reaction plane with the projectile nucleus moving along the $+z$ direction at $x=b/2$ and the target nucleus moving along the $-z$ direction with $x=-b/2$.  Then the angular momentum of the produced matter is oriented in the $-y$ direction.  The baryon species are assumed to all couple to the vector meson with similar coefficients.  (See the discussion below.)  Therefore we approximate
\be
m^2_{V} V^{\mu} = \bar{g}_{V} J_{\rm B}^{\mu}
\ee
with an effective coupling $\bar{g}_{V}$.  We write the baryon current as
$J_{\rm B}^0 = n_{\rm B}(t)$ and $\vec{J}_{\rm B} = n_{\rm B}(t) \vec{v}(\vec{x},t)$ with the velocity parametrized by
\be
\vec{v} = \left( \dot{\psi}_x(t) x + {\rm c}_1 z/t, \dot{\psi}_y(t) y, z/t + {\rm c}_3 x/t \right) \,.
\ee  
The third component with $z/t = \tanh\eta$, where $\eta$ is space-time rapidity, is the usual longitudinal expansion in the Bjorken model \cite{Bjorken}.  The $\dot{\psi}_x(t) x$ and $\dot{\psi}_y(t) y$ terms represent transverse expansion and when they are different they reflect elliptic flow.  The ${\rm c}_1$ term represents directed flow of the baryons as they are deflected away from the beam axis.  The ${\rm c}_3$ term represents shear flow along the beam axis.  The ${\rm c}_i$ terms represent contributions to vorticity since $\vec{\nabla}\times\vec{v} = (0, \Delta {\rm c}/t, 0)$, where $\Delta {\rm c} = {\rm c}_1 - {\rm c}_3$, which can be positive or negative.  Baryon conservation leads to $\dot{n}_{\rm B}(t) + (\dot{\psi}_x(t) + \dot{\psi}_y(t) +t^{-1}) n_{\rm B}(t) = 0$.  In general, for fixed transverse coordinate one expects the $\dot{\psi}$ to start near zero, rise with time and then fall to zero.  Since we are interested in the time around hadronization we take $\dot{\psi}_x(t) = a_x/t$ and $\dot{\psi}_y(t) = a_y/t$ with $a_x$ and $a_y$ constants.  Then
\be
{n}_{\rm B}(t) = {n}_{\rm B}(t_{\rm ch}) \left(\frac{t_{\rm ch}}{t}\right)^{a_x + a_y +1}
\ee
where $t_{\rm ch}$ is the time of hadronization.  The limit $a_x = a_y = 0$ corresponds to longitudinal expansion only, while the limit $a_x = a_y = 1$ corresponds to homologous spherical expansion.  Consistent with this is the approximation that the scalar density $n_s$ is a function of $t$ only so that the scalar field does not contribute to the polarization, at least in this simple model.

This is basically a blast wave model.\footnote{One may change variables so that $x = \tau \sinh \rho \cos \phi$ and $x/t = (\sinh \rho/\cosh \eta) \cos \phi$ where $\rho \ge 0$ is the transverse rapidity, and similarly for $y$.}  At some time $t_{\rm f} > t_{\rm ch}$ hydrodynamic flow ceases and free-streaming begins.  At that time  $x^2+y^2 \lesssim R^2$, where $R$ is a cutoff on the transverse extent of the matter.  The resulting fields are
\ba
\vec{B}_{V} &=& \frac{\bar{g}_{V}}{m^2_{V}} \left(\frac{\Delta {\rm c}}{t}\right) n_{\rm B}(t) \, \hat{\vec{y}} \nonumber \\
\vec{E}_{V} &=& \frac{2\bar{g}_{V}}{m^2_{V}} \left(\frac{1+a_{\rm av}}{t}\right) n_{\rm B}(t) \, \vec{v}
\ea
with
\be
\vec{\nabla}\times  \vec{E}_V = -\frac{\partial \vec{B}_V}{\partial t} = 
\frac{2\bar{g}_{V}}{m^2_{V}} \left(\frac{1+a_{\rm av}}{t^2}\right) n_{\rm B}(t) \Delta {\rm c} \, \hat{\vec{y}}
\ee
where $a_{\rm av} \equiv \thalf (a_x+a_y)$.  Then the ratio of the second to the first coefficient in Eq. (\ref{e2}) is $(1+a_{\rm av})/(2 m_\Lambda t)$ which is less than 6\% for $ t > 3$ fm/c.  Due to the symmetries of our simple model, $\langle \vec{E}_V \times \vec{p} \rangle = 0$ when averaging is done with a Boltzmann distribution boosted by the flow velocity $\vec{v}$, so the third term in Eq. (\ref{e2}) is also negligible.  The second and third terms do not contribute to the polarization difference anyway.

Suppose that the spins were in equilibrium at temperature $T(t_{\rm ch})$ at time $t_{\rm ch}$.  With the quantization axis in the $y$ direction the average polarization in the 
high-temperature/weak-field limit would be
\be
P_y =  \beta \frac{g_{V\Lambda}}{m_{\Lambda}} \frac{|\vec{B}_{V}|}{2T} 
 = g_{V\Lambda} \bar{g}_{V} \frac{n_{\rm B}(t_{\rm ch})}{m_{\Lambda} m_V^2} \frac{\Delta {\rm c} \, \beta}{2 t_{\rm ch} T(t_{\rm ch})} \,.
\label{result1}
\ee
Assuming that $g_{V\Lambda} \bar{g}_{V} > 0$ this implies that $\Lambda$'s are polarized in the $+y$ direction while $\bar{\Lambda}$'s are polarized in the $-y$ direction if $\Delta {\rm c} > 0$, and the opposite if $\Delta {\rm c} < 0$.  

Synergy with neutron star physics comes from the requirement that the Lagrangian produce a relativistic mean field equation of state that is stiff enough to support stars of at least two solar masses \cite{Demorest,Antoniadis}.  Introduction of new degrees of freedom, such as hyperons, softens the equation of state and lowers the maximum mass \cite{JEllis}.  This provides a constraint on the values of the vector coupling constants so that the appearance of hyperons is delayed to higher densities.  For example, Ref. \cite{SU3} used SU(3) flavor symmetry along with ideal $\omega$-$\phi$ mixing to determine the vector couplings in terms of a singlet $g_1$, an octet $g_8$, and the F/(F+D) ratio $\alpha$.  They are
\ba
g_{\omega N} &=& \oneth (4\alpha - 1) g_8 + \sqrt{\twoth} g_1 \nonumber \\
g_{\omega \Lambda} &=& - \twoth (1 - \alpha) g_8 + \sqrt{\twoth} g_1 \nonumber \\
g_{\omega \Sigma} &=& \twoth (1 - \alpha) g_8 + \sqrt{\twoth} g_1 \nonumber \\
g_{\omega \Xi} &=& - \oneth (2\alpha + 1) g_8 + \sqrt{\twoth} g_1 \nonumber \\
g_{\phi N} &=& \textstyle{\frac{\sqrt{2}}{3}}(4\alpha - 1) g_8 - \sqrt{\oneth} g_1 \nonumber \\
g_{\phi \Lambda} &=& - \textstyle{\frac{2\sqrt{2}}{3}} (1 - \alpha) g_8 - \sqrt{\oneth} g_1 \nonumber \\
g_{\phi \Sigma} &=& \textstyle{\frac{2\sqrt{2}}{3}} (1 - \alpha) g_8 - \sqrt{\oneth} g_1 \nonumber \\
g_{\phi \Xi} &=& - \textstyle{\frac{\sqrt{2}}{3}} (2\alpha + 1) g_8 - \sqrt{\oneth} g_1 \,.
\ea
Examples of numerical values which produce a two solar mass neutron star are given in Table I.  Those choices were used for illustration in Ref. \cite{SU3} because the SU(6) values are $\alpha=1$ and $g_8/g_1=1/\sqrt{6}$.  Greater masses are produced when $g_8/g_1 < 0.35$ for $\alpha=1$ and when $\alpha < 0.9$ for $g_8/g_1=1/\sqrt{6}$.
\begin{table}[thp]
\centering
{\renewcommand{\arraystretch}{1.3}
\begin{tabular}[t]{| c| c | c | } 
 \hline
Vector  & $\alpha=1$ &  $\alpha=0.9$  \\
Coupling  & $g_8/g_1=0.35$ & $g_8/g_1=1/\sqrt{6}$  \\
\hline
$g_{\omega\Lambda}/g_{\omega N}$  & 0.700  & 0.674  \\ 
$g_{\omega\Sigma}/g_{\omega N}$  & 0.700  & 0.721  \\ 
$g_{\omega\Xi}/g_{\omega N}$  & 0.400 & 0.372 \\
\hline
$g_{\phi N}/g_{\omega N}$  & -0.071 & -0.066 \\
$g_{\phi\Lambda}/g_{\omega N}$  & -0.707  & -0.526  \\ 
$g_{\phi\Sigma}/g_{\omega N}$  & -0.495  & -0.460  \\
$g_{\phi\Xi}/g_{\omega N}$  & -0.919 & -0.954 \\ 
 \hline
\end{tabular}
}
\caption{Examples of SU(3) couplings that produce a two solar mass neutron star \cite{SU3}}.
\end{table}

Synergy also arises with hypernuclear physics \cite{Gal2016}.  A survey of results in the literature for $\Lambda$ binding energies in nuclei leads to $g_{\omega \Lambda} \approx 0.55 g_{\omega N}$ and $g_{\sigma \Lambda} \approx 0.5 g_{\sigma N}$ \cite{Kapusta-Gale}.  Unlike nucleons, the spin-orbit interaction experienced by the $\Lambda$ is an order of magnitude smaller than that arising from the Foldy-Wouthuysen reduction of the vector interaction.   The physical reason for this is that, within the quark model, the spin of the $\Lambda$ is carried by the strange quark which does not couple to the $\omega$ meson \cite{Jennings1,Cohen}.  To cancel the spin-orbit interaction arising from the vector interaction requires the tensor interaction
\be
{\cal L}_{\rm tensor} = - \frac{f_{\omega\Lambda}}{4m_{\Lambda}} \bar{\psi}_{\Lambda} \sigma^{\mu\nu} \psi_{\Lambda} \omega_{\mu\nu} \,.
\ee
This contributes to the spin-orbit energy an amount
\be
H^{\rm tensor}_{\rm spin-orbit} = \frac{f_{\omega\Lambda}}{m^2_\Lambda}
\frac{1}{r}\frac{\partial \omega_0}{\partial r} \vec{S} \cdot \vec{L} \,.
\label{TSO}
\ee
Cancellation with the vector and scalar couplings implies that $f_{\omega\Lambda} \approx - g_{\omega\Lambda}$.  The Zeeman energy
\bd
- \frac{(g_{\omega\Lambda}+f_{\omega\Lambda}) }{m_\Lambda} \beta \ \vec{S} \cdot \vec{B}_{\omega} 
\ed
is thus suppressed.  This suggests that it is the $\phi$ meson which is the primary origin of the polarization difference, which should be no surprise because of the matching of the strange quark spin content of these hadrons.

In general there is no reason to expect that vector and tensor interactions between all vector mesons and all baryons will cancel.  For example, quark models have been used to estimate that $(f/g)_{\omega\Sigma}$ ranges between 0.6 and 1.3, and that $(f/g)_{\omega\Xi}$ ranges between $-1.9$ and $-2.3$ \cite{Cohen}.  Reference \cite{Jennings2} used the values 1 and $-2$, respectively.  There is no empirical knowledge of the spin-orbit coupling involving the $\phi$ meson due to the lack of observation of multiply strange hypernuclei \cite{Gal2016,privateGal}.

There are several complications before one is able to make precision comparisons to data.  These include, but are not limited to, feed-down from decays of the heavier hyperons $\Sigma$, $\Xi$, and $\Omega$; feedback of the polarized spins to produce an effective vector meson magnetic field via susceptibility; and a more realistic, relativistic space-time evolution of the baryon current. Nevertheless, we make some preliminary comparisons here.  The difference in polarization in the $-y$ direction according to Eq. (\ref{result1}) has the form
\be
P_{\bar{\Lambda}} - P_{\Lambda} = C \left( \frac{n_{\rm B}(t_{\rm ch})}{0.15/{\rm fm}^3} \right) \left( \frac{140 \, {\rm MeV}}{T(t_{\rm ch})} \right) \,.
\ee
For the chemical potential and temperature at $t_{\rm ch}$ as functions of $\sqrt{s_{NN}}$ we use the parametrization given in \cite{freezeoutcurve}.  We then use a crossover equation of state from \cite{matchingpaper} to determine the baryon density.  For illustration, since the precise magnitude is rather uncertain for the reasons given above, we consider two cases.  In case I $C$ is independent of beam energy.  In case II $C \sim 1/\sqrt{s_{NN}}$ because generally the directed flow and the shear flow of net baryons is expected to decrease with increasing energy.  We take $C=0.03$ for case I and  $C = 0.45 \, {\rm GeV}/\sqrt{s_{NN}}$ for case II; both assume that $\Delta {\rm c} > 0$.  The coefficients are chosen to give a reasonable visual fit to the polarization data as shown in Fig. 1.  The difference in polarizations rises with decreasing energy because the net baryon density increases, the temperature decreases, and in case II the factor $C$ rises with decreasing energy.  It is interesting to note that the directed flow of both net protons \cite{directedSTAR1} and net $\Lambda$'s \cite{directedSTAR2} is actually negative in the range $10 < \sqrt{s_{NN}} < 30$ GeV.  This may reflect a change in the equation of state of the produced matter \cite{Nara}.  Because the polarization difference is sensitive to the baryon current it is a probe of the reaction dynamics.
\begin{figure}[h]
\includegraphics[width=0.99\columnwidth]{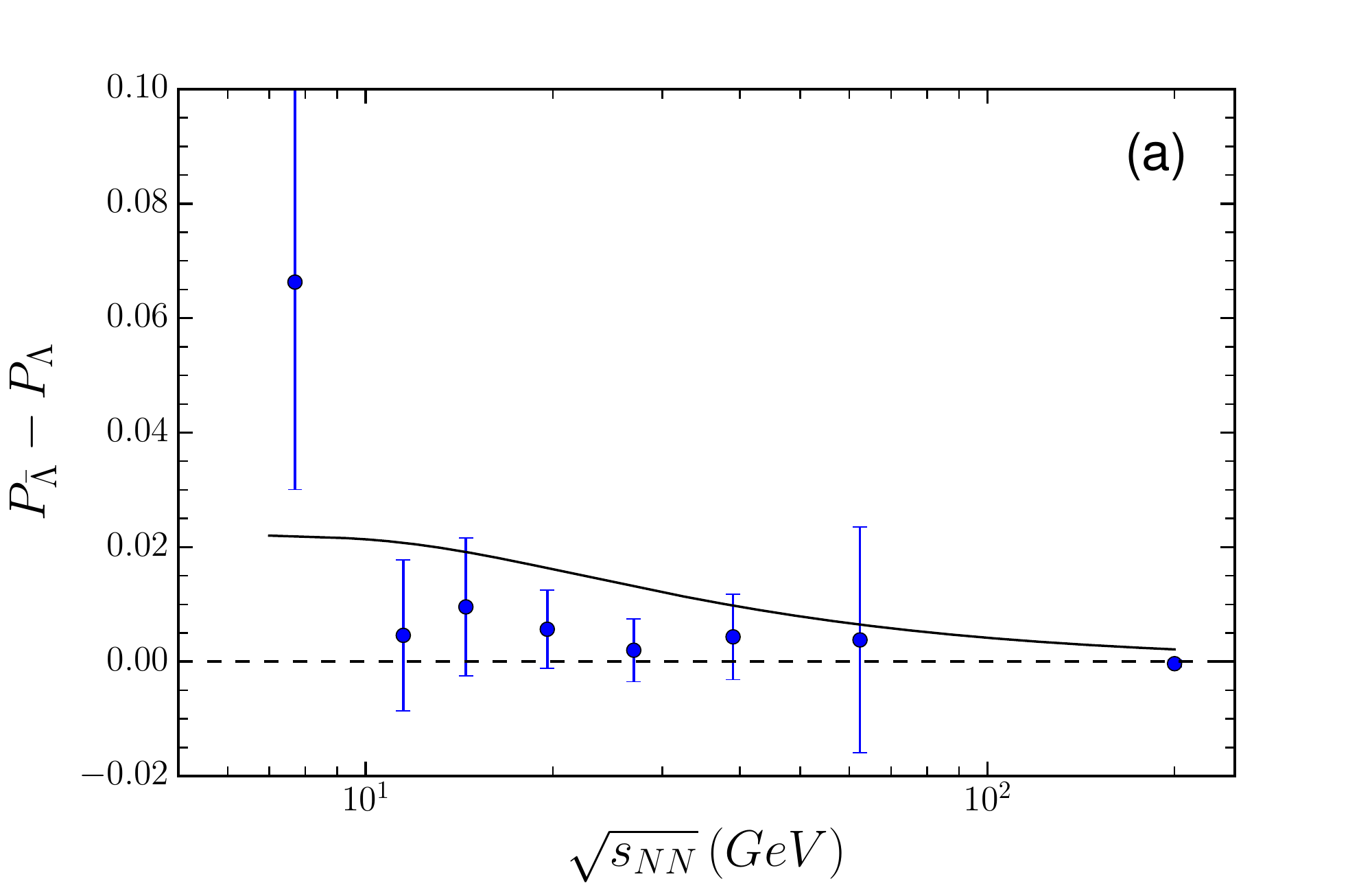}
\includegraphics[width=0.99\columnwidth]{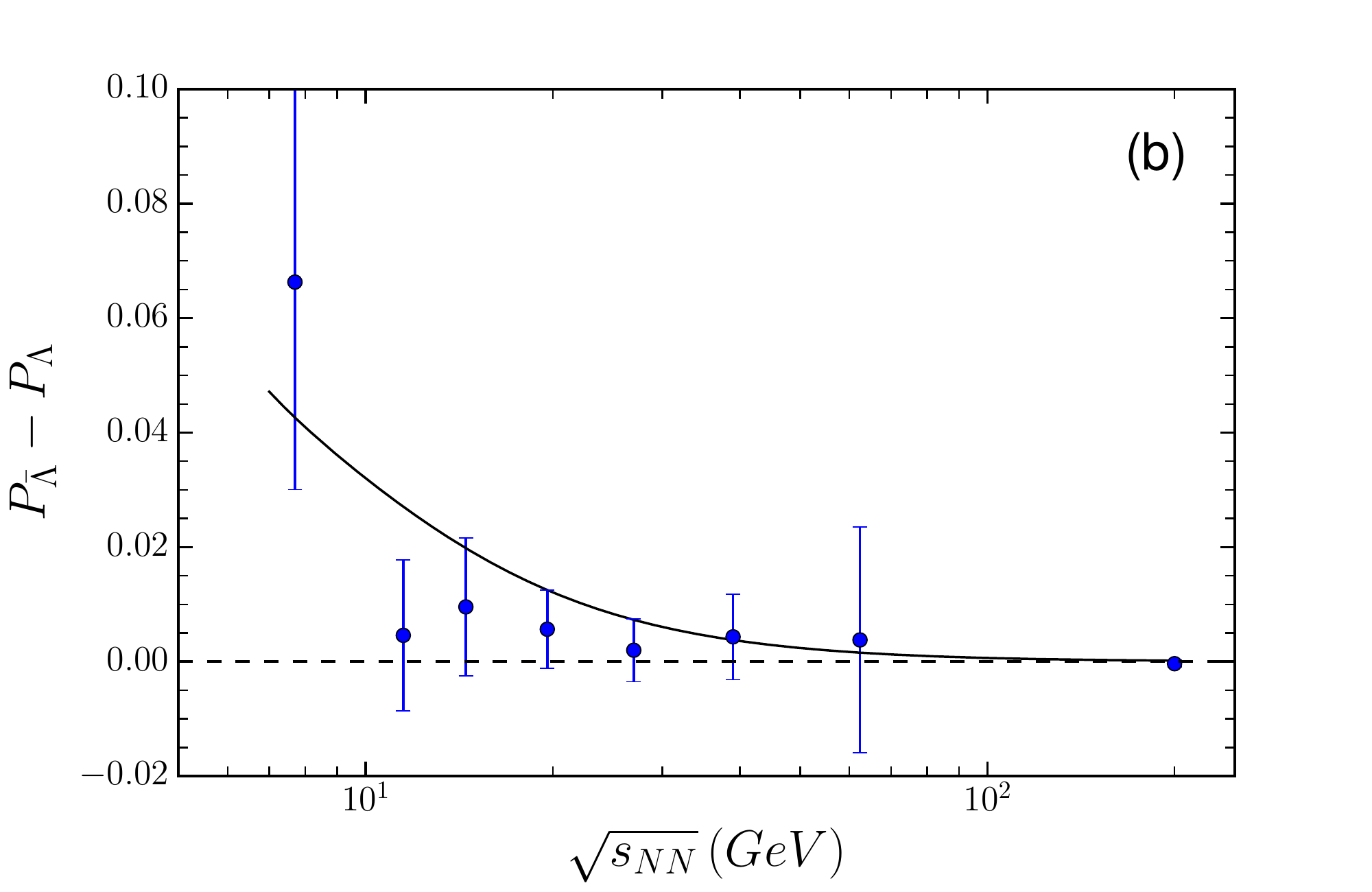}
\caption{(color online) Difference in polarization of $\Lambda$ and $\bar{\Lambda}$ hyperons, with positive value meaning that it is opposite to the total angular momentum of the produced matter.  The top panel (a) shows case I and the bottom panel (b) shows case II, as described in the text.  The data at 200 GeV come from \cite{SecondSTAR}, the rest come from \cite{Nature}.  Only statistical uncertainties are included.}
\label{fig:1}
\end{figure}

For comparison the true magnetic field produced in high energy heavy ion collisions points in the $-y$ direction.  The equilibrium $\Lambda$ polarization due to that field is $P_y =-\mu_{\Lambda} B/T$ which orients the spin in the $+y$ direction because the magnetic moment is negative: $\mu_{\Lambda} = - 0.61 \mu_N$ where $\mu_N$ is the nuclear Bohr magneton.  Being its antiparticle, the $\bar{\Lambda}$ would be polarized in the $-y$ direction.  The magnetic field has been calculated with the inclusion of the electrical conductivity $\sigma_E$ of the produced matter; in its absence the magnetic field at the time of hadronization is orders of magnitude smaller \cite{Tuchin}.  At time $t$ at $z=0$ its value is
\be
B = \frac{e b \sigma_E}{8\pi t^2} \exp(- b^2 \sigma_E/4t)
\ee
where $b$ is the impact parameter.  Evaluated at $t=t_{\rm ch} = 3$ fm/c, $b=7$ fm, $T=T(t_{\rm ch}) = 140$ MeV, and $\sigma_E = 6$ MeV the magnitude of the polarization is
$|P_y| =7.4\times 10^{-6}$, totally irrelevant compared to the strong interaction induced polarization.  Note also that as long as the condition $\gamma_{\rm beam} b \sigma_E > 1$ is satisfied there is no beam energy dependence to the magnetic field.  Realistic transport model calculations show that the time extent of the magnetic field is on the order of 0.2 fm/c, which is too short to build up observable polarization \cite{Voronyuk}.

For the problem of relaxation of a small departure from equilibrium we turn to studies in the area of spintronics.  A solution to the Bloch equations for a static magnetic field in the $y$ direction provides a formula for the spin relaxation rate $\Gamma_s$ for the magnetization in that direction in the form \cite{Fabian,Szolnoki}
\be
\Gamma_s = \frac{\langle \Omega_x^2\rangle + \langle \Omega_z^2\rangle}{\Omega_y^2 + \Gamma_c^2} \Gamma_c \,.
\ee
Here $\Omega_y$ is the Larmor frequency associated with the static magnetic field, $\langle \Omega_x^2\rangle$ and $\langle \Omega_z^2\rangle$ are the average fluctuations of the Larmor frequencies in the perpendicular directions, and $1/\Gamma_c$ is the coherence time of a single spin, which we take to be the time between scatterings of the hyperons with other particles.  The fluctuations in this problem arise from the spin-orbit term involving $\vec{E}_\omega \times \vec{p}$ in Eq. (\ref{e2}).  We apply this formula assuming an adiabatic evolution of the vector meson magnetic field in the $y$ direction.  Around $t_{\rm ch}$ the time between collisions is on the order of several fm/c, so that $\Omega_y < \Gamma_c$.  From the spin-orbit interaction we estimate that $\Gamma_s \ll \Gamma_c$, and therefore the polarization difference should be established around the time of hadronization at its equlibrium value and should not change significantly thereafter.  

In conclusion, we have argued that well-known interactions of baryons with mesons can result in a splitting of the polarizations of $\Lambda$ and $\bar{\Lambda}$ hyperons in high energy heavy ion collisions.  This interaction is orders of magnitude larger than the one arising from electromagnetic fields. The results are sensitive to the space-time evolution of the baryon current.  There is much work to be done by theorists and experimentalists to fully exploit this idea.   

\section*{Acknowledgments}
We thank A. Gal, J. W. Halley, A. Kamenev, M. Li, D. J. Millener, and C. Plumberg for discussions. The work of L. Cs. was supported by the Research Council of Norway Grant No. 255253/F50 and the work of J. K. and T. W. was supported by the U.S. DOE Grant No. DE-FG02-87ER40328.



\begin{thebibliography}{99}

\bibitem{QMseries}
See the proceedings of the Quark Matter Conference series, the most recently available being ``The 26th International Conference on Ultra-relativistic Nucleus-Nucleus Collisions: Quark
Matter 2017", edited by U. Heinz, O. Evdokimov, and P. Jacobs [Nucl. Phys. A {\bf 967}, (2017)].

\bibitem{whitepaper}
First Three Years of Operation of RHIC [Nucl. Phys. A {\bf 757}, Issues 1-2 (2005)].

\bibitem{Wang1}
Z. T. Liang and X. N. Wang, Phys. Rev. Lett. {\bf 94}, 102301 (2005), [Erratum: Phys. Rev. Lett. {\bf 96}, 039901 (2006)].

\bibitem{Becattini1}
F. Becattini, F. Piccinini, and J. Rizzo, Phys. Rev. C {\bf 77}, 024906 (2008).

\bibitem{FirstSTAR}
B. I. Abelev {\it et al.} (STAR Collaboration),  Phys. Rev. C {\bf 76}, 024915 (2007), [Erratum: Phys. Rev. C {\bf 95}, 039906 (2017)].

\bibitem{Nature}
L. Adamczyk {\it et al.} (STAR Collaboration), Nature (London) {\bf 548}, 62 (2017).

\bibitem{SecondSTAR}
J. Adam {\it et al}. (STAR Collaboration), Phys. Rev. C {\bf 98}, 014910 (2018).

\bibitem{Becattini2}
F. Becattini, L. P. Csernai, and D. J. Wang, Phys. Rev. C {\bf 88}, 034905 (2013).

\bibitem{Becattini3}
F. Becattini, I. Karpenko, M. Lisa, I. Upsal, and S. Voloshin, Phys. Rev. C {\bf 95}, 054902 (2017).

\bibitem{nuclearSpinOrbit}
M. Goeppert-Mayer, Phys. Rev. {\bf 75}, 1969 (1949); Haxel, Jensen, and Suess, {\it ibid}. {\bf 75}, 1766 (1949).

\bibitem{Duerr}
H.-P. Duerr, Phys. Rev. {\bf 103}, 469 (1956).

\bibitem{Walecka74}
J. D. Walecka, Ann. Phys. (N.Y.) {\bf 83}, 491 (1974). 

\bibitem{Serot-Walecka}
B. D. Serot and J. D. Walecka, Adv. Nucl. Phys. {\bf 16}, 1 (1986); B. D. Serot, Rep. Prog. Phys. {\bf 55}, 1855 (1992); 
B. D. Serot and J. D. Walecka, Int. J. Mod. Phys. E {\bf 6}, 515 (1997). 

\bibitem{Kapusta-Gale}
J. I. Kapusta and C. Gale, {\it Finite Temperature Field Theory} (Cambridge University Press, Cambridge, 2006).

\bibitem{Dover-Gal}
C. B. Dover and A. Gal, Prog. Part. Nucl. Phys. {\bf 12},  171 (1984).

\bibitem{FW}
L. L. Foldy and S. A. Wouthuysen, Phys. Rev. {\bf 78}, 29 (1950).

\bibitem{BD}
J. D. Bjorken and S. D. Drell, {\it Relativistic Quantum Mechanics} (McGraw-Hill, New York, 1964).

\bibitem{Greiner}
W. Greiner, {\it Relativistic Quantum Mechanics} (Springer-Verlag, Berlin, Germany, 1987).

\bibitem{Hinschberger}
Y. Hinschberger and P.-A. Hervieux, Phys. Lett. A {\bf 376}, 813 (2012).

\bibitem{Hickey}
M. C. Hickey and J. S. Moodera, Phys. Rev. Lett. {\bf 102}, 137601 (2009).

\bibitem{Mondal}
R. Mondal, M. Berritts, and P. M. Oppeneer, Phys. Rev. B {\bf 94}, 144419 (2016);
R. Mondal, M. Berritts, A. K. Nandy, and P. M. Oppeneer, {\it ibid}. {\bf 96}, 024425 (2017).

\bibitem{Bjorken}
J. D. Bjorken, Phys. Rev. D {\bf 27}, 140 (1983).

\bibitem{Demorest}
P. B. Demorest, T. Pennucci, S. M. Ransom, M. S. E. Roberts, and J. W. T. Hessels, Nature (London) {\bf 467} 1081 (2010). 

\bibitem{Antoniadis}
J. Antoniadis {\it et al.}, Science {\bf 340}, 6131 (2013).

\bibitem{JEllis}
J. Ellis, J. I. Kapusta, and K. A. Olive, Nucl. Phys. B {\bf 348}, 345 (1991).

\bibitem{SU3}
S. Weissenborn, D. Chatterjee, and J. Schaffner-Bielich, Phys. Rev. C {\bf 85}, 065802 (2012).

\bibitem{Gal2016}
A. Gal, E. V. Hungerford, and D. J. Millener, Rev. Mod. Phys. {\bf 88}, 035004 (2016).

\bibitem{Jennings1}
B. K. Jennings, Phys. Lett. B {\bf 246}, 325 (1990).

\bibitem{Cohen}
J. Cohen and H. J. Weber, Phys. Rev. C {\bf 44}, 1181 (1991).

\bibitem{Jennings2}
J. Mares and B. K. Jennings, Phys. Rev. C {\bf 49}, 2472 (1994).

\bibitem{privateGal}
A. Gal (private communication).

\bibitem{freezeoutcurve} 
J. Cleymans, H. Oeschler, K. Redlich, and S. Wheaton, Phys. Rev. C {\bf 73}, 034905 (2006).

\bibitem{matchingpaper} 
M. Albright, J. Kapusta, and C. Young, Phys. Rev. C {\bf 90}, 024915 (2014); {\bf 92}, 044904 (2015).

\bibitem{directedSTAR1}
L. Adamczyk {\it et al}. (STAR Collaboration), Phys. Rev. Lett. {\bf 112}, 162301 (2014).

\bibitem{directedSTAR2}
L. Adamczyk {\it et al}. (STAR Collaboration),  Phys. Rev. Lett. {\bf 120}, 62301 (2018).

\bibitem{Nara}
Y. Nara, H. Niemi, A. Ohnishi, J. Steinheimer, X. Luo, and H. St\"ocker, Eur. Phys. J. A {\bf 54}, 18 (2018).

\bibitem{Tuchin}
K. Tuchin, Phys. Rev. C {\bf 88}, 024911 (2013).

\bibitem{Voronyuk}
V. Voronyuk, V. D. Toneev, W. Cassing, E. L. Bratkovskaya, V. P. Konchakovski, and S. A. Voloshin, Phys. Rev. C {\bf 83}, 054911 (2011).

\bibitem{Fabian}
J. Fabian, A. Matos-Abiague, C. Ertler, P. Stano, and I. Zutic, Acta Physica Slovaca {\bf 57}, 565 (2007).  See especially Eq. (IV.36).

\bibitem{Szolnoki}
L. Szolnoki, B. D\'ora, A. Kiss, J. Fabian, and F. Simon, Phys. Rev. B {\bf 96}, 245123 (2017).


\end{thebibliography}
\end{document}